\documentclass[prd, aps, superscriptaddress, preprintnumbers, twocolumn, floatfix, nofootinbib]{revtex4-2}
\pdfoutput=1

\usepackage{amsfonts}
\usepackage{amsmath}
\usepackage{amssymb}
\usepackage{bm}
\usepackage{dcolumn}
\usepackage{graphicx}
\usepackage[latin1]{inputenc}
\usepackage{latexsym}
\usepackage{rotating}
\usepackage{hyperref}
\usepackage{graphicx}
\usepackage{xcolor}

\newcommand\be{\begin{equation}}
\newcommand\bea{\begin{eqnarray}}
\newcommand\ee{\end{equation}}
\newcommand\eea{\end{eqnarray}}

\renewcommand{\d}{{\mathrm{d}}}

\renewcommand{\]}{\right]}
\renewcommand{\(}{\left(}
\renewcommand{\)}{\right)}

\newcommand{\Mpl}{M_{\textrm{Pl}}}

\def\doi{http://doi.org}

\def\d{\mathrm{d}}

\begin{document}

\title{Emergent Metric Space-Time from Matrix Theory}

\author{Suddhasattwa Brahma}
\email{suddhasattwa.brahma@gmail.com}
\affiliation{Higgs Centre for Theoretical Physics, School of Physics \& Astronomy,
University of Edinburgh, Edinburgh EH9 3FD, UK.}

\author{Robert Brandenberger}
\email{rhb@physics.mcgill.ca}
\affiliation{Department of Physics, McGill University, Montr\'{e}al, QC, H3A 2T8, Canada}

\author{Samuel Laliberte}
\email{samuel.laliberte@mail.mcgill.ca}
\affiliation{Department of Physics, McGill University, Montr\'{e}al, QC, H3A 2T8, Canada}

\date{\today}

%%%%%%%%%%%%%%%%%%%%%%%%%%%%%%%%%%%%%%%%%%%%%%%%%%%%%%%%%%%%%%%%%%%%%%%%%%%%%%%%%%%%%%%%%%%%%%
\begin{abstract} 
The IKKT matrix model yields an emergent space-time. We further develop these ideas and give a proposal for an emergent metric. Based on previous numerical studies of this model, we provide evidence that the emergent space-time is continuous and infinite in extent, both in space and in time, and that the metric is spatially flat. The time evolution describes the transition from a string-theoretic emergent phase to a phase in which the $SO(9)$ symmetry of the model is spontaneously broken to $SO(6) \times SO(3)$, with three dimensions of space expanding, becoming classical and at later times evolving like in a radiation-dominated universe, and the remaining six dimensions of space stabilized at the string scale.  We speculate on how this analysis can be extended to yield an early universe cosmology which, in addition to the above-mentioned properties, also leads to a roughly scale-invariant spectrum of cosmological fluctuations and gravitational waves. 
\end{abstract}
%%%%%%%%%%%%%%%%%%%%%%%%%%%%%%%%%%%%%%%%%%%%%%%%%%%%%%%%%%%%%%%%%%%%%%%%%%%%%%%%%%%%%%%%%%%%%%

\maketitle

\section{Introduction}
Evidence is mounting that in order to obtain a model of the very early universe which is consistent with the required quantum treatment of matter, we must go beyond a description based on naive effective field theory \footnote{By {\it naive} effective field theory we mean the usual approach to  (Fock)-quantizing fields in which the fields are expanded into comoving modes, and each such mode is canonically quantized like the standard quantization of a harmonic oscillator.}. One of the reasons is based on unitarity problems of any effective field theory description of an expanding universe \cite{Weiss, AS}. A ultraviolet cutoff is required to make sense of such a model, and to maintain this ultraviolet cutoff at a fixed physical scale in an expanding universe, continuous creation of effective field theory modes is required. This  time-dependence of the low-energy  Hilbert space obviously implies the breakdown of unitarity. Demanding that, during the time evolution of the system, no such modes which are created ever exit the Hubble horizon (when these modes freeze out, become squeezed and can classicalize) leads to the {\it Trans-Planckian Censorship Conjecture} (TCC) \cite{TCC1} which results in serious constraints on inflationary cosmology \cite{TCC2} (see e.g. \cite{TCCrev} for recent discussions).

Another problem for an effective field theory analysis of cosmology is the following:
Since each of the modes of the effective field theory has a ground state energy, such an effective field theory analysis leads to the famous cosmological constant problem, the fact that the predicted value of the vacuum energy is many orders of magnitude larger than what is consistent with observations  if the vacuum energy gravitates.

Keeping these problems in mind, it is worth considering that although the {\it inflationary scenario} \cite{Guth} has become the standard paradigm for early universe cosmology, there are alternative scenarios which are also consistent with current data. One class involves {\it bouncing cosmologies} (see e.g. \cite{bouncerev} for reviews) in which the universe is initially contracting and then undergoes a bounce transition to an expanding phase, while another class is the {\it emergent scenario} (see e.g. \cite{SGCrev} for a review), in which the universe begins in a quasi-static phase and then undergoes a phase transition to an expanding radiation phase. To realize bouncing and emergent cosmologies, however, physics beyond standard effective field theory of matter and Einstein gravity is required. 

To obtain a consistent description of the early universe, we hence need to start with a model which is well defined and has no problems at high energy scales. For a long time, the hope has been that superstring theory will be able to provide such a description. The phenomenology of string theory is, however, usually explored in an effective field theory limit, and in this limit the aforementioned conceptual problems cannot be resolved.  Thus, instead of trying to evaluate four-dimensional effective potentials using string-theoretic quantum effects, we believe a more fruitful way would be to start with the full theory itself and give a prescription for coarse-graining the cosmological degrees of freedom from it.

In this paper we provide some indications that a viable emergent cosmology will emerge from matrix model  descriptions of string theory. Specifically, we consider the IKKT matrix model \cite{IKKT}, a proposed non-perturbative definition of Type IIB superstring theory. This is a quantum mechanical model of Hermitean $N \times N$ matrices (with, a priori,  no space and no time). We will indicate how in the $N \rightarrow \infty$ limit a continuous space-time with three large spatial dimensions and with infinite extent both in space and time emerges. We provide a prescription for an emergent metric for the $(3 + 1)$-dimensional space-time involving the three large spatial dimensions, and provide indications that the emergent metric is spatially flat, and described by a cosmological scale factor $a(t)$ which has a late time limit which corresponds to a radiation dominated universe.  The effective cosmological constant vanishes in this model. \footnote{The late time analysis does not take into account the presence of fermionic matrices,  and matter will arise from that sector.}.

In the following we first review the IKKT matrix model \cite{IKKT}. Over the past two decades there has been a lot of numerical work on this model (see e.g. \cite{IKKTrev, Ydri} for reviews, and \cite{IKKTnew} for more recent numerical studies), and we will summarize the results which are relevant for our analysis. In Section 3, we then show how continuous space-time with infinite extent of both space and time variables emerges, give a proposal for an emergent metric, and show that the resulting cosmological metric corresponds to a spatially flat manifold. In Section 4, we speculate that the same result will also emerge in the BFSS matrix model, a matrix theory (with an intrinsic time, but no space) which was proposed \cite{BFSS} as a non-perturbative definition of M-theory\footnote{For other related approaches to emergent space in the context of matrix models see e.g. \cite{Jevicki},  \cite{DVV}, \cite{CSV}, \cite{BMN}, \cite{Smolin}.  For early work on matrix models as a means to quantize a theory of membranes see \cite{Hoppe}.}. As shown in previous work \cite{us}, thermal fluctuations in a high temperature state of the BFSS model yield scale-invariant spectra of curvature fluctuations and gravitational waves, with a Poisson component of the curvature perturbations on short distance scales.

We will be working with units in terms of which the speed of light, Boltzmann's constant and Planck's constant are all set to $1$.

\section{Review of the IKKT Matrix Model and Emergence of Continuous Time}

The IKKT matrix model \cite{IKKT} (see \cite{IKKTrev, Ydri} for recent reviews) has been proposed as a non-perturbative definition of Type IIB superstring theory. It is a pure matrix theory (no space and no time), given by the action
\be \label{IKKTaction}
	S \, = \,  -\frac{1}{g^2} \text{Tr}\(\frac{1}{4} \[A^a, A^b\]\[A_a,A_b\] + \frac{i}{2} \bar{\psi}_\alpha \({\cal{C}} \Gamma^a\)_{\alpha\beta} \[A_a,\psi_\beta\]\)\,,
\ee
where $A_a$ and $\psi_\alpha$ ($a=0,\ldots,9$, $\alpha =1,\ldots,16$, ) are $N\times N$ are bosonic and fermionic Hermitian matrices, respectively, the $\Gamma^{\alpha}$ are the gamma-matrices for $D = 10$ dimensions, and ${\cal{C}}$ is the charge conjugation matrix. Note that $a$ is a ten-dimensional vector index, while $\alpha$ is a spinor index. The vector indices are raised and lowered with the Minkowski  symbol $\eta_{ab}$. $g$ is a gauge-theory coupling constant. It is in the limit $N \rightarrow \infty$ with $\lambda \equiv g^2 N$ held fixed that this action leads to a non-perturbative definition of Type IIB superstring theory. 

The action of the Lorentzian matrix model \cite{IKKT} is given by the following functional integral over the bosonic and fermionic matrices (with the standard measures)
\be
Z \, = \, \int dA d\psi e^{iS} \, .
\ee
 
Since the matrices are Hermitean, it is possible to choose a basis in which $A_0$ is diagonal. We can also label the basis elements such that the eigenvalues $\alpha_a$ are ordered such that $\alpha_a < \alpha_b$ if $a < b$. Numerical studies of the theory show that for large values of $N$ \cite{IKKTnum1}
\be \label{tempres}
 \frac{1}{N} \left\langle {\rm Tr} A_0^2 \right\rangle \,  \sim \, \kappa N \, ,
 \ee
 where $\kappa$ is a constant ($\kappa < 1$), where the pointed  brackets in $\left\langle {\cal{O}}\right\rangle$ indicate the expectation value of the operator ${\cal{O}}$
 \be \label{evalue}
 \left\langle {\cal{O}} \right\rangle  \, \equiv \, \frac{1}{Z} \int dA\, d\psi\, {\cal{O}}\, e^{iS} \, .
 \ee
 This implies that in the $N \rightarrow \infty$ limit, the total extent of time becomes infinite, time running from $ - \infty$ to $+ \infty$. More precisely, time runs from $-t_{m}$ to $+t_{m}$ with $t_m$ scaling as $\sqrt{N}$. To see this, assume for concreteness that the temporal eigenvalues are evenly spaced, with spacing $\Delta t$. In this case, the expectation value on the left hand side of  (\ref{tempres}) becomes the sum of squares of integers from $1$ to $N/2$, multiplied by $(\Delta t)^2$. Making use of the formula for the sum of squares of integers, we find that the left hand side of (\ref{tempres}) is proportional to $N^2 (\Delta t)^2$. Hence, time becomes continuous with 
 \be \label{Deltat}
 \Delta t \, \sim \frac{1}{\sqrt{N}} \, ,
\ee
and the total extent of space becomes infinite,  as $N \rightarrow \infty$
\be
t_m \, \sim \, N \Delta t \, \sim \, \sqrt{N} \, .
\ee

Turning our attention now to the spatial matrices $A_i$, numerical work \cite{IKKTnum1} has shown that these matrices have band-diagonal structure in the sense that if we consider the expectation values of the off-diagonal elements of the matrix, then they decay to zero if the distance $n$ from the diagonal exceeds a critical value $n_c$, \textit{i.e.}
\be \label{decay}
\sum_i \left\langle | A_i |_{ab}^2 \right\rangle \, \rightarrow \, 0 \,\,\, \rm{for} \,\,\, n \equiv |a - b| > n_c \, .
\ee
The numerical studies \cite{IKKTnum1} also show that
\be
n_c \, \sim \, \sqrt{N} \, .
\ee
This result can also be understood by first performing the partial functional integral $dA_{ab}$ in (\ref{evalue}).  Schematically,   
\begin{eqnarray}\label{R1}
	\left\langle |A_i|_{ab}^2 \right\rangle \sim \int d | A_i |_{ab}\,  \dfrac{| A_i |_{ab}^2}{Z} \, e^{i/2g^2 \left(\alpha_a -\alpha_b\right)^2 | A_i |_{ab}^2}	
\end{eqnarray}
where
\bea
Z = \int \prod_a  d\alpha_a \prod_{a>b}\, \left(\alpha_a -\alpha_b\right)^2 \, \int d A_i e^{i/2g^2 \left(\alpha_a -\alpha_b\right)^2 | A_i |_{ab}^2}
\eea
For values of $|A_i|_{ab}^2 \left(\alpha_a - \alpha_b\right)^2/g^2 >1$, \textit{i.e.} when $|a - b| > n_c$, the integrand becomes very rapidly oscillating, and by the Riemann-Lebesgue Lemma the integral hence tends to zero (in the sense of generalized  functions). In the following we will assume that
\be \label{constant}
\sum_i  \left\langle | A_i |_{ab}^2 \right\rangle \, \sim \, {\rm{const}} \,\,\, \rm{for} \,\,\, n \equiv |a - b| < n_c \, ,
\ee
which is supported by the same consideration of the $dA_{ab}$ integral.  From the above \eqref{R1}, it is also easy to check that the scale $n_c$ scales as (for $a-b =n_c$):
\bea
\frac{\left(\alpha_a - \alpha_b\right)^2}{g^2} \sim 1 \, \Rightarrow  n_c \sim g\,\sqrt{N}\,,  
\eea
where we have used the relation $\alpha_a - \alpha_b = n_c \Delta t \sim n_c/\sqrt{N}$ (from \eqref{Deltat}).

Let us briefly return to the temporal matrix $A_0$. A  time variable $t(m)$ corresponding to the m'th temporal eigenvalue can then be defined by averaging the diagonal elements $\alpha_i$ over $n$ elements \cite{IKKTtime}
\be
t(m) \, \equiv \, \frac{1}{n} \sum_{l = 1}^n \alpha_{m + l} \, ,
\ee
Time-dependent spatial matrices $({\bar{A_i}})_{I, J}(t)$ of dimension $n \times n$ can then be defined via \cite{IKKTspace}
\be
({\bar{A_i}})_{I, J}(t(m)) \, \equiv \, (A_i)_{m + I, m + J} \, .
\ee
It is then natural to define the extent $x_i$ of a given spatial dimension $i$ at time $t$ by
\cite{IKKTextent}
\be \label{extent}
R_i(t)^2 \, \equiv \, \left\langle \frac{1}{n} \text{Tr} ({\bar{A_i}})(t)^2 \right\rangle \, ,
\ee
where the pointed brackets stand for the quantum expectation value in the state given by the partition function.  

\begin{figure}
\includegraphics[scale = 0.7]{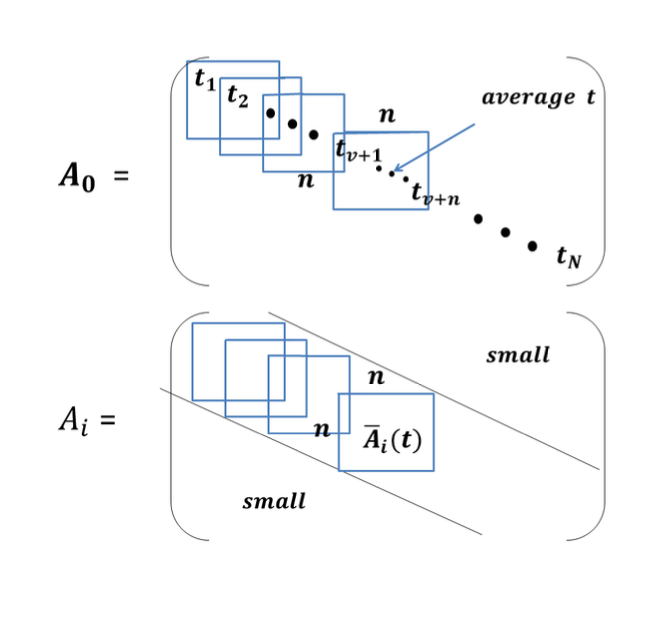}
\caption{Temporal matrix $A \equiv A_0$ (top panel) and spatial matrices $A_i$ (bottom panel)  in the basis in which the temperal matrix is diagonal. The spatial matrices $A_i$ (bottom panel) have ``block-diagonal form'' and can be used to define the sizes of the spatial dimensions at time $t$ via sub-matrices ${\bar{A_i}}(t)$ of $A_i$ centered a ``distance'' $t$ along the diagonal of $A_i$. This figure is taken from \cite{Ydri} with permission.} 
\label{matrix}
\end{figure}

Various numerical studies \cite{IKKTpt} of the IKKT mode indicate the as $t$ increases, the initial $SO(9)$ symmetry of the matrix model is spontaneously broken to $SO(6) \times SO(3)$, with three of the $R_i(t)$ increasing in time in a nearly isotropic manner, while the other six remain small. If we identify $R_i(t)$ as the {\it extent of space parameter} in direction $i$ as a function of tiime (which is done in the work on the IKKT matrix model), then we find the emergence of precisely three large spatial dimensions. Using Gaussian expansion analyses, one can verify that the state with $SO(6) \times SO(3)$ symmetry has a smaller free energy than the state with the full $SO(9)$ symmetry, and that this symmetry breaking pattern is in fact preferred over other ones.

This symmetry breaking which leads to three large spatial dimensions with the other six remaining small occurs in {\it String Gas Cosmology} \cite{BV}. In that model, a gas of closed strings in considered on a nine dimensional spatial torus. At high densities, the string gas will contain, in additional to the center of mass momentum modes and the string oscillatory modes, winding modes -- strings winding the torus. It is then argued that for space to be able to expand, the winding modes need to annihilate into string loops, and this cannot happen in more than 3 large spatial dimensions since for more than three large spatial dimensions the strings would have vanishing probability to meet. We conjecture that there might be a relationship between the symmetry breaking dynamics in the matrix model and in String Gas Cosmology: there is evidence that the strings are coherent states of the matrix model in which one spatial matrix is excited. Such a coherent state will prevent space from expanding in the same way that a string does in the String Gas Cosmology picture. 

\section{Emergent Space and Metric}

Let us now return to the spatial matrices $A_i$ for the dimensions which become large. As already indicated in the previous section, we now consider a $n_i \times n_i$ submatrix ${\bar{A}}_i^{n_i}(t)$ centered a distance $t$ down the diagonal, as indicated in Figure \ref{matrix}. We let $n_i$ range from $0$ to $n_c$, and we propose to view $n_i$ as a comoving spatial coordinate in direction $i$ \footnote{The comoving coordinate thus defined runs from $0$ to $n_c$. With a slight change of the definition we can extend the range to $-n_c < n_i < n_c$. We focus on the upper (lower) diagonal submatrix of the $|n_i| \times |n_i|$ matrix for positive (negative) values of $n_i$. The resulting length (and metric) will be symmetric under $n_i \rightarrow -n_i$.}. Then, we adapt (\ref{extent}) and define the physical length of a curve along the $i$ coordinate axis from $n = 0$ to $n$ as
 \be \label{phys}
 l_{i, {\rm{phys}}}^2(t, n_i) \, \equiv \, \left\langle \text{Tr} ({\bar{A_i}^{n_i}})(t))^2 \right\rangle \, .
 \ee

Making use of (\ref{constant}), we see that the quantity (\ref{phys}) scales as $n_i^2$ (there are $n_i$ eigenvalues to sum over, and each eigenvalue obtains contributions from $n_i$ matrix elements). Thus, the total physical extent of space out to a comoving distance $n_c$ scales as $n_c \sim N^{1/2}$, and the physical distance $\delta x$ between neighboring comoving coordinate values scales as 
\be
\delta x \, \sim \, \frac{N^{1/2}}{N} \, \sim \, N^{-1/2} \, . 
\ee
Thus, we obtain continuous space with infinite spatial extent in the $N \rightarrow \infty$ limit.

Since for a metric space the physical length of a line along the $i$ axis between comoving coordinates $0$ and $x$ allows us to obtain the $g_{ii}$ metric component via
\be
 l_{i, {\rm{phys}}}(t, x) \, = \int_{y = 0}^x \sqrt{g_{ii}(t, y)} dy \, ,
 \ee
%%, 
we propose to define the emergent metric as
\be
g_{ii}^{1/2}(t, n_i) \, = \, \frac{d}{dn_i} l_{i, {\rm{phys}}}(t, n_i) \, .
\ee
Making use of (\ref{phys}) we get
\be
g_{ii}^{1/2}(t, n_i) \, = \, 
\frac{1}{2} \dfrac{\left( \frac{d}{dn_i} \left\langle \text{Tr} \left({\bar{A_i}^{n_i}}\right)^2(t) \right\rangle \right)}{\left( \left\langle \text{Tr} \left({\bar{A_i}^{n_i}}\right)^2(t) \right\rangle \right)^{1/2}} 
 \, .
\ee
Since (based on (\ref{constant}) we have seen that (\ref{phys}) scales as $n_i^2$, we find that the metric component is independent of $n_i$.  This is a very important finding as it tells us that the emergent spacetime is spatially flat. Note that this crucially depends on the block diagonal structure of the spatial matrices with respect to our coarse-grained time. If, indeed, the spatial matrices were all diagonal or of some other form, then the scaling of the metric of $n_i^2$ would have been very different and would not have led to this result. The block-diagonal form, in turn, depends sensitively on the Lagrangian of the IKKT model and this type of an emergent structure would not appear in any matrix model but rather one that comes from string theory.

In the end, we obtain
\be
g_{ii}(n_i, t) \, = \, {\cal{A}}(t) \delta_{ii} \,  ,
\ee
where ${\cal{A}}(t)$ is the time-dependent amplitude. This result corresponds to a homogeneous and commutative space for the three large dimensions. Making use of the $SO(3)$ symmetry of the system we then obtain a homogeneus and isotropic cosmological metric
\be
g_{ij}(t) \, = \, {\cal{A}}(t) \delta_{ij} \, .
\ee

We identify the amplitude ${\cal{A}}(t)$ with the cosmological scale factor $a(t)$ of the three-dimensional space which becomes large. At early times (before the symmetry breaking phase transition) the scale factor is constant and corresponds to a state of string density. After the phase transition, the scale factor increases. Since quantum effects are expected to become negligible once the amplitude ${\cal{A}}(t)$ is large (by Ehrenfest's Theorem), the  late time dependence of ${\cal{A}}(t)$ can be obtained by solving the classical equations of motion. This has been done in \cite{Nish} with the result that ${\cal{A}}(t) \sim t^{1/2}$ which corresponds to the expansion of space in the radiation phase of Standard Big Bang cosmology.

We have thus obtained a first principles realization of the String Gas Cosmology (SGC) scenario put forwards in \cite{BV}. The SGC scenario is obtained by considering matter to be a thermal gas of strings on a nine-dimensional background space which admits long-lived winding string states. There is a maximal temperature of a gas of closed strings, the Hagedorn temperature \cite{Hagedorn} $T_H$. For a large range of energy densities, the temperature $T$ remains close to $T_H$, and it is not unreasonable to assume that this phase (the {\it Hagedorn phase}) is quasi-static. At some point, however, a symmetry breaking phase transition occurs \cite{BV}, allowing three dimensions of space to become large. This transition is triggered by the decay of winding modes. When the world sheet of two winding strings with opposite orientations meet, they can interact and produce string loops, thus eliminating the winding which prevents space from expanding, and leading to a radiation-dominated phase of expansion. Since string world sheets have zero probability for intersecting in more than four large space-time dimensions, it is precisely three spatial dimensions which can become large, while the others remain at the string scale. The weak point of the SGC scenario is the assumption of a quasi-static Hagedorn phase. Such a phase cannot be obtained using effective field theory techniques. 

We have argued that matrix theory can provide a first principles realization of the dynamics of space-time assumed in SGC. The universe begins in a quasi-static phase in which all nine spatial extent parameters are the same, time independent and of string scale. This corresponds to the Hagedorn phase of SGC. There is a phase transition in which the $SO(9)$ symmetry of the Lagrangian is spontaneously broken to $SO(6) \times SO(3)$. The same transition also occurs in SGC. After the phase transition, the three dimensions which become large expand like in the radiation phase of Standard Big Bang cosmology, the same dynamics as once again occurs in SGC. 

Note that there are other approaches to obtaining SGC dynamics from first principles. In the approach of \cite{Vafa}, the analog of the Hagedorn phase of SGC is a topological phase. In the context of Double Field Theory \cite{DFT}, there have also been recent studies on how to obtain an initial cosmological phase which has properties in common with the Hagedorn phase of SGC \cite{Franz}.

Note that, as already pointed out in \cite{CC}, the cosmological constant problem is absent in this model. The quantization of the model does not involve an effective field theory analysis in which fields are expanded in Fourier modes, and the ground state energy of the Fourier modes then adds up to yield a cosmological constant which is many orders of magnitude too large. Here, we are quantizing a matrix model Lagrangian, extracting an effective late time metric, and observing that the time evolution is inconsistent with the presence of a cosmological constant. We thus see that the cosmological constant problem may be an artefact of an effective field theory point of view.

\section{Discussion}

At about the same time that the IKKT matrix model was proposed, there was another proposal for a non-perturbative definition of string theory, namely the BFSS \cite{BFSS} matrix model. In contrast to the IKKT model, this model contains a time variable $t$, and is given by a Lagrangian involving $9$ spatial Hermitean $N \times N$ matrices $X_i(t)$ and a temporal Hermitean $N \times N$ matrix $A_0(t)$:
\bea \label{BFSSaction}
S \, &=& \, \frac{1}{2 g^2} \int dt \bigl[ {\rm Tr} \left \{\frac{1}{2} (D_t X_i)^2 - \frac{1}{4} [X_i, X_j]^2\right\} \nonumber \\
& & - \theta^T D_t \theta - \theta^T \gamma_i [ \theta, X^i ] \bigr] \, ,
\eea
where  $D_t \equiv \partial_t - i [A_0(t), ..]$ is a covariant derivative operator. The $\theta$ are sixteen fermionic superpartners which are spinors of $SO(9)$, and $g^2$ is a coupling constant. In the large $N$ limit, (with the `t Hooft coupling $\lambda \equiv g^2 N = g_s N l_s^{-3}$ (where $g_s$ and $l_s$ are string coupling constant and string length, respectively) held fixed, this model was argued to yield a non-perturbative definition of M-theory.

In the high temperature limit, the leading term in the action of the bosonic sector of the BFSS matrix yields the bosonic part of the IKKT action. This is seen by expanding the $X_i$ matrices in terms of Matsubara frequencies
\be
X_i (t) \, = \, \sum_{m=0}^{\infty} X_i^m e^{i m \omega t} \, ,
\ee
with $\omega = 2\pi / \beta$, $\beta$ being the inverse of the temperature $T$, and $m$ running over the positive semi-definite integers. The BFSS action becomes
\be
S \, = \, S_0 + S_1 \, ,
\ee
where $S_0$ contains the terms which only depend on the zero modes $X_i^0$. At high temperature, the terms in $S_1$ are small amplitude correction terms, and under the rescaling
\be
A_i \, \equiv \, T^{1/4} X_i^0 \, ,
\ee
the action $S_0$ becomes the bosonic part of the IKKT action.

Starting with the BFSS model in a high temperature thermal state, we recently showed (using results from \cite{Kawahara}) that the thermal fluctuations which are inevitably present in such a state lead to scale-invariant spectra of curvature fluctuations and gravitational waves, with a Poisson tail which dominates in the ultraviolet in the case of the curvature fluctuations \cite{us}. The results parallel those obtained in the case of String Gas Cosmology \cite{NBV, BNPV}, in which case the scale-invariance can be traced back to the fact that the fluctuations have holographic scaling as a function of the radius of the box in which the fluctuations are computed. Note that it is the correction terms $S_1$ in the high temperature expansion of the BFSS action which play an important role in determining the fluctuations.

We now suggest the following scenario: starting from the BFSS matrix model and taking a high temperature thermal state, we extract an emergent space-time and an emergent metric using the results from the IKKT model presented above. In order to be able to apply these arguments, we need to make sure that the $SO(9) \rightarrow SO(6) \times SO(3)$ symmetry breaking also occurs in the case of the BFSS model. Since the fermions appear to play an important role in the phase transition in the IKKT model \cite{IKKTpt}, this is a non-trivial assumption. However, we have recently shown \cite{us2} that a phase transition which breaks the $SO(9)$ symmetry indeed occurs in the BFFS model \cite{us2}. Next, we also need to verify that the evolution of the expectation values $\left\langle {\rm{Tr}} | A_0 |^2 \right\rangle$ and $\left\langle | A_i |_{ab}^2 \right\rangle$ is not changed. If successful, we would have a first principles realization of an emergent early universe cosmology in which infinite range continuous time, infinite range continuous space (with exactly three large spatial dimensions) and a homogeneous and spatially flat metric which leads to a radiation-dominated three dimensional expanding space all naturally emerge from the matrix model. Thermal fluctuations then lead to scale-invariant spectra of density fluctuations and gravitational waves with an ampitude which is set by $(\eta_s / m_{pl})^4$, where $\eta_s$ is the string energy scale, and $m_{pl}$ is the Planck mass \cite{us}.

In the appendix, we give a simple toy model realization of late-time dynamics for the BFSS action. Instead of going through the steps mentioned above, we use the classical equations of motion of the BFSS model to find cosmological solutions using a time gauge that had been overlooked earlier to find a universe with radiation-dominated expansion. If nothing else, this serves as an indication that a realization of a ultraviolet-complete model, as above, maybe possible in the BFSS model.

\section{Conclusions}

We have reviewed how continuous and infinte range time and space emerge from the IKKT matrix model in the $N \rightarrow \infty$ limit. As has been shown, this model undergoes symmetry breaking between an early stage of $SO(9)$ symmetry in which the {\it extent of space parameters} in all nine spatial directions are microscopic, and a stage when the $SO(9)$ symmetry breaks to $SO(6) \times SO(3)$, allowing exactly three spatial dimensions to become large. We have proposed a definition of comoving distance coordinates and corresponding physical distance, with which it becomes possible to extract an emergent metric for the four large space-time dimensions. The resulting metric is homogeneous, isotropic and spatially flat. At late times, the three large dimensions expand like in the radiation-dominated Friedmann model. Hence, the matrix model leads to vanishing cosmological constant, as already pointed out in \cite{CC} \footnote{For other approaches towards extracting cosmology from the IKKT matrix model see \cite{Klinkhamer} and \cite{Steinacker}.}.

We have suggested that the method of extracting time, space and a metric from a matrix model also holds if we start from the BFSS matrix theory. Based on this starting point, we also have a mechanism by which thermal fluctuations lead to scale-invariant spectra of curvature fluctuations and gravitational waves.

\section*{Acknowledgments}

The authors thank Simon Caron-Huot,  Keshav Dasgupta, Hikaru Kawai, Jun Nishimura, Katsuta Sakai, Lee Smolin and Cumrun Vafa for comments on an earlier version of this draft. We also thank B. Ydri for permission to use the figure (\ref{matrix}) which is taken from \cite{Ydri}.

SB is supported in part by the Higgs Fellowship. SL is supported in part by FRQNT. The research at McGill is supported in part by funds from NSERC and from the Canada Research Chair program.

\appendix

\section*{Appendix: Late-time evolution in the BFSS model}
Although we have described how one can extract a metric for the IKKT model in this paper, let us show some evidence that, at late times, one does find a cosmological evolution for the BFSS model on the lines of what we have conjectured above. Our main assumptions in deriving such dynamics are as follows:
\begin{enumerate}
	\item We can examine the classical equations of motion of the BFSS model in order to gain some insight into late-time evolution. We expect that this assumption certainly breaks down before the critical time when we expect the spontaneous symmetry-breaking to take place. 
	\item We assume a \textit{homothetic anstaz}, as have been previously chosen in \cite{Classical_BFSS}.
\end{enumerate}

The classical equations for the bosonic matrices from the action \eqref{BFSSaction} is given by
\begin{eqnarray}\label{EoM_BFSS}
	\ddot{X}^i + \left[X^j, \left[X^j, X^i\right]\right] = 0 \,,
\end{eqnarray}
where repeated indices are summed over irrespective of their position. Since we are assuming these equations to hold only after the symmetry-breaking takes place, it is natural to suppose that the $i$ index now runs from $(1,2,3)$. The homothetic ansatz implies 
\begin{eqnarray}\label{Homo_ansatz}
	X^i = a(\tau)\, \Theta^i\,.
\end{eqnarray}
Before solving \eqref{EoM_BFSS} with our ansatz above \eqref{Homo_ansatz}, let us make a few remarks. Firstly, note that the BFSS model implies classical equations for a Galilei invariant system. In doing so, we depart from previous studies where it was customary to add a mass term for the matrices which led to the symmetry group being that of a Newton-Hooke system (the non-relativistic contraction of de Sitter). Not only would adding such a mass term be unnatural from the point of view of the BFSS model (as derived from M-theory), requiring a positive cosmological constant would, in fact, imply having a negative (or tachyonic) mass term. 

Secondly, note that the time parameter is intrinsic in the BFSS model unlike its IKKT counterpart. Therefore, it is not clear how to relate the time parameter appearing in these equations with the different choices of time gauges one can choose in General Relativity. However, this is where we make our most important conjecture -- what if the equation of motion written above \textit{describes dynamics with respect to conformal time?} Although, at this point, this is simply a hypothesis on our part, note that this choice is just as good as choosing the time parameter to be equivalent to cosmic time, as has been typically assumed in the past. In future work, we will explore more into this question of which time gauge does this time parameter correspond to for our cosmological scenario. For now, we simply identify with proper time and go on to explore the consequences. 

Finally, as in well known for the BFSS model, the dynamics described by \eqref{EoM_BFSS} is topped off by the following (Gauss) constraint on initial conditions:
\begin{eqnarray}
	\left[\dot{X}^i, X^i\right] = 0\,.
\end{eqnarray}

Since all the time-dependence of our ansatz \eqref{Homo_ansatz} is contained in the overall pulsating function, which we have suggestively called $a(\tau)$ to coincide with the scale factor, \eqref{EoM_BFSS} can be written as
\begin{eqnarray}
	\frac{\ddot{a}(\tau)}{a^3(\tau)} + \left[\Theta^j, \left[\Theta^j, \Theta^i\right]\right] \,\(\Theta^i\)^{-1} = 0\,,
\end{eqnarray}
where both the terms on the left hand side must be independent of time now. (Keep in mind that an overdot denotes a derivative with respect to conformal time in our convention.) Let us choose the constant for the separation of variables to be $\lambda$, \textit{i.e.}
\begin{eqnarray}
	\ddot{a}(\tau) &=& \lambda \, a^3(\tau)\,,\label{EOM1}\\
	\left[\Theta^j, \left[\Theta^j, \Theta^i\right]\right] &=& -\lambda \Theta^i \,.\label{EOM2}
\end{eqnarray}
Let us begin analyzing the equation \eqref{EOM1} as the Raychaudhuri equation for the scale factor, whose first integral gives the more familiar Friedmann equation:
\begin{eqnarray}
	\frac{d}{d\tau} \left(\dot{a}(\tau)\right) &=& \lambda a^3(\tau) \Rightarrow \frac{\dot{a}^2(\tau)}{2} = \dfrac{\lambda a^4(\tau)}{4} + K\,,
\end{eqnarray}
where $K$ is a constant of integration. Next, we can switch over to cosmic time, recalling $d/d\tau = a(t) d/\d t$, and doing a little bit of algebra to find
\begin{eqnarray}\label{Friedmann1}
	\frac{1}{a^2} \left(\dfrac{d a}{d t}\right)^2 = \dfrac{K}{a^4} + \dfrac{\lambda}{2}\,.
\end{eqnarray}
This is the standard Friedmann equation with a radiation component as well as a cosmological constant term (the sign of $\lambda$ is yet to be determined).

However, recall how we expect these classical equations to only be valid for late times in which regime we expect the bosonic matrices to commute with each other since they are far separated and give rise to ordinary smooth spacetime geometry. In this case, for near-commuting matrix degrees of freedom, it is natural to have $\lambda = 0$ as can be seen by inspecting \eqref{EOM2}. If we  plug it into the above Friedmann equation \eqref{Friedmann1}, we get
\begin{eqnarray}\label{Friedmann}
	3 \Mpl^2	H^2 = \dfrac{C}{a^4}\,,
\end{eqnarray}
where $C = 3\Mpl^2 K$ is the constant for the radiation energy density. In fact, this is where the fact that we consider that this equation is satisfied after the symmetry-breaking phase comes into play. For a $9$-d universe, the equation of state corresponding to the Friedmann equation \eqref{Friedmann} would be given by $p = - (5/9)\; \rho$. However, for $d=3$, this is given by $ p=  \rho/3$, as expected. We will come back to this point later on.

We could have also set $\lambda = 0$ directly into the Raychaudhuri equation \eqref{EOM1} to find $\ddot{a} = 0$, whose solutions are given by
\begin{eqnarray}
	a(\tau) = a_0 + a_1 \tau\,.
\end{eqnarray}
As expected, this is indeed the solution for the scale factor for a radiation dominated universe along with a possible quasi-static phase. It is interesting that this is indeed the type of cosmological history predicted by the String Gas scenario (a pressureless fluid corresponding to the quasistatic phase followed by radiation)! Of course, we should only trust this solution after the symmetry-breaking phase and only find the description of the emergent (or non-geometric) phase as quasistatic to be intriguing. 

Let us compare this result with previous investigations of BFSS cosmology \cite{Classical_BFSS}. Since it was always assumed that the time parameter corresponds to cosmic time in existing literature, this inevitably led to unphysical matter components along with a possible curvature term. Sometimes, a mass term was added by hand to the BFSS Lagrangian which led to a cosmological constant term. Apart from the radiation phase, what is remarkable is that using our interpretation of the time parameter as conformal time, one immediately finds that there is \textit{no spatial curvature term} in the Friedmann equation above! This is consistent with our finding of a flat metric from the IKKT model as described in the previous section. Moreover, this conclusion is completely independent of our natural supposition that the matrices become commuting at late times. Even if we choose to keep a non-zero $\lambda$, this would act as an effective cosmological constant and not behave as a curvature term in the Friedmann equation. This is one of our main findings from the BFSS model -- unless there are additional terms put in by hand, the Friedmann equations describe a flat, radiation dominated universe at late time with our choice of the time variable.

Finally, let us comment about choosing to work with $d=3$ in interpreting the above equations. At first sight, this might seem like an extremely restrictive condition and the reader might view this as our most drastic assumption. However, this is not the case. To begin with, as explained, it only makes sense to use the classical equations to describe the dynamics at late times and this would naturally be after the symmetry-breaking phase. So, the real question is whether there is any such symmetry-breaking phase in the BFSS model, and this has recently been shown to be the case \cite{us2}. But, more importantly, note that if we allow for a parametric separation between the magnitude of the scale-factor $a(t)$ of the external cosmological background and those of the $6$-dimensional internal spacetime, say $\tilde{a}(t)$, then it is break-up the original equations of motion \eqref{EoM_BFSS} into two parts -- one for the external and one for the internal spacetimes. Given this, it is perfectly possible that the internal spacetimes \textit{do not} become commuting at late time and indeed has a $\tilde{\lambda}$ that is nonzero. That would simply imply that the Friedmann equation for the internal scale factor $\tilde{a}$, has an additional cosmological constant term in $\tilde{\lambda}$. In fact, as has been shown in \cite{Classical_BFSS}, this constant is typically negative and takes the value $\tilde{\lambda} =-2$ if we assume the internal dimensions to have an $SO(3) \times SO(3)$ form. For an $SO(6)$ symmetry, $\lambda$ would also be negative but a little more complicated to evaluate. What is important is that in deriving our Friedmann equation \eqref{Friedmann} above, we do not need to assume a BFSS  toy model for $d=3$. All we need is the knowledge that the solution of the classical BFSS equations of motion would be satisfied by an anisotropic ansatz since the scale factors of the internal and external dimensions must be parametrically separated due to the symmetry-breaking phase preceding it. However, we are still working with the full BFSS model, as derived from M-theory, for arriving at our solution for the $(3+1)$-d cosmological scale factor.

\end{document}